\documentstyle[11pt,titlepage]{article}

\newcommand{\be}{\begin{equation}}
\newcommand{\ee}{\end{equation}}

\newcommand{\bea}{\begin{eqnarray}}
\newcommand{\ena}{\end{eqnarray}}

\newcommand{\ba}{\begin{array}}
\newcommand{\ea}{\end{array}}

\newcommand{\ibar}{{\bar{\imath}}}

\begin{document}

\begin{titlepage}
\begin{center}
{\LARGE {\bf Gaugino condensation and the anomalous $U(1)$}} 
\vskip 1cm

{\large P. Bin\'etruy}\\
{\em LPTHE\footnote{Laboratoire associ\'e au CNRS-URA-D0063.},
Universit\'e Paris-XI, B\^atiment 211,  F-91405 Orsay Cedex, France} 
\\[.7cm] 

{\large E. Dudas} \\
{\em CEA-SACLAY, Service de Physique Th\'eorique} \\
{\em F-91191 Gif-sur-Yvette Cedex, France} \\[1.2cm]

\end{center} 

\centerline{ {\bf Abstract}}

\indent
We study gaugino condensation in presence of an anomalous $U(1)$
gauge group and find that global supersymmetry is dynamically broken.
An example of particular interest is provided by effective string
models with 4-dimensional Green-Schwarz anomaly cancellation
mechanism. The structure of the hidden sector is
constrained by the anomaly cancellation conditions and the scale of
gaugino condensation is shifted compared with the usual case. We
explicitly compute  the resulting soft supersymmetry
breaking terms.

\vfill \rightline{LPTHE-Orsay \ \ 96/60}  
\rightline{Saclay T96/075} 
\rightline{hep-ph/9607172}

\end{titlepage}
\section{Introduction}
Among the scenarios for breaking supersymmetry, the condensation 
of gauginos in a hidden sector remains a favourite one \cite{Nilles}. 
Gaugino condensation is indeed central to the idea of dynamical
supersymmetry breaking in the context of gauge symmetries. But in its
explicit realizations in superstring models \cite{DRSW}, it suffers
from a number of drawbacks. The degeneracies  associated with the flat
directions of the scalar potential are lifted but the corresponding
degrees of freedom such as the dilaton are not stabilized and the true
ground state is found  at infinite field values where supersymmetry is
restored. In order to overcome this problem, so-called ``racetrack''
or multicondensate models  \cite{racetrack} have been  proposed where
two terms of different orders conspire in order to stabilise the field
and provide a supersymmetry-breaking minimum.

Many of the superstring models have an anomalous $U(1)_X$
gauge symmetry \cite{DSW} which could play a important role in issues
such as fermion mass hierarchies \cite{IR,ramond}, cosmology
\cite{cosmo} and ... gaugino condensation. The latter connection was
recently stressed by Banks and Dine \cite{BD}.\footnote{ We wish to
thank Luis Ib\'{a}\~{n}ez for drawing our attention to this paper and
the whole issue.} Indeed, the anomalous $U(1)_X$ has mixed anomalies
with the other gauge symmetries --those of the standard model as well
as the hidden sector--, anomalies which are cancelled through a
4-dimensional Green-Schwarz mechanism \cite{GS} using the couplings of
the dilaton superfield to the gauge superfields. It is therefore not
surprising that the whole issue of supersymmetry breaking through
gaugino condensation is deeply modifed in such models. Moreover,
because in the Green-Schwarz mechanism all the mixed anomalies are
non-vanishing and proportional to one another, there must exist fields
charged under $U(1)_X$ in the observable as well as in the hidden
sector. The $U(1)_X$ gauge symmetry thus serves as a messenger
interaction competitive with the gravitational interaction. In this
paper, we wish to stress the modifications that the presence of such
an anomalous $U(1)_X$ symmetry is bringing to the scenario of a
dynamical supersymmetry breaking through gaugino condensation.

To be more specific, the relevant couplings of the dilaton superfield
$S$ to the gauge superfields $V_a, V_X$ (of respective gauge
invariant field strengths $W^{\alpha}_a, W^{\alpha}_X$) of the
groups $G_a$ and $U(1)_X$ read in the  global limit:
\begin{eqnarray}
{\cal L}_{S,V} &=& - \int d^4 \theta \ln (S+ S^+ -
\delta_{GS}V_X) \nonumber \\
& & +   \int d^2\theta \left[{S \over 4} (\sum_a k_a{\rm Tr}
W^{\alpha}_a W_{a\alpha} + k_X {\rm Tr} W^{\alpha}_X W_{X\alpha}) +
{\rm h.c.}\right] \label{eq:L} \end{eqnarray}
where $\delta_{GS}$ is the Green-Schwarz coefficient and $k_a$
($k_X$) is the Kac-Moody level of the group $G_a$ ($U(1)_X$).
Under a $U(1)_X$ gauge transformation ($A^X_\mu \rightarrow A^X_\mu
+ \partial_\mu \alpha$),
$S$ is shifted as 
\begin{equation}
S \rightarrow S + {i \over 2} \delta_{GS} \alpha(x). \label{eq:shift}
\end{equation}
The complete Lagrangian is invariant provided the mixed
$U(1)_X[G_a]^2$ anomaly coefficients $C_a$ satisfy the condition
\begin{equation}
\delta_{GS} = {C_a \over k_a} = {C_X \over k_X} = {C_g \over k_g},
\label{eq:Ck} \end{equation}
where $C_g$ is the mixed gravitational anomaly proportional to
${\rm Tr} X$. Indeed a string computation yields  
\begin{equation}
\delta_{GS} = {1 \over 192 \pi^2} {\rm Tr} X.
\label{eq:deltaGS}
\end{equation}
The mixing between $S$ and $V_X$ in the K\"ahler potential (\ref{eq:L})
gives rise to a D-term in the scalar potential:
\begin{equation}
V_D = {g^2_X \over 2} \left(\sum_A X_A K_A \phi^A + {1 \over 4}
k_X g^2_X \delta_{GS}   M^2_P \right)^2, \label{eq:VD}
\end{equation}
where $M_P$ is the Planck scale,
\begin{equation}
k_X g^2_X = {2 \over S+S^+},
\end{equation}
and $K_A$ is the derivative of the K\"ahler potential $K$ with respect
to the field $\phi^A$.  The presence of the Fayet-Iliopoulos term
induced by the $U(1)_X$ anomaly usually induces a non-zero vacuum
expectation value for one (or more) field of $X$ charge of sign
opposite to $\delta_{GS}$.

In the case where there is no anomalous $U(1)$, a non-anomalous
R-symmetry under which the $S$ superfield undergoes a translation
similar to (\ref{eq:shift}) imposes that the scale $\Lambda$ at which
the hidden sector gauge coupling becomes strong (which sets the scale
for the corresponding gaugino condensates) behaves as
\begin{equation}
\Lambda \sim M_P e^{-{kS \over 2 b_0}}, \label{eq:Lambda}
\end{equation}
where $b_0$ is the one-loop beta function coefficient for the hidden
sector gauge group and $k$ is its Kac-Moody level. This is obviously
not invariant under the $U(1)_X$ transformation (\ref{eq:shift}) of
$S$. This shows that  the presence of an anomalous $U(1)_X$ necessarily
modifies the standard discussion of gaugino condensation.

This study goes beyond the case of
effective superstring models. Indeed, it is not unusual that, in the
course of sequential gauge symmetry breaking, appears an anomalous
$U(1)_X$ abelian gauge symmetry whose mixed anomalies are cancelled
through a Green-Schwarz type mechanism using an effective degree of
freedom with dilaton-axion couplings. Thus we will consider in what
follows the general case of a gauge model with symmetry group $SU(N_c)
\times U(1)_X$ with a non-vanishing Fayet-Iliopoulos term and the
following matter content: $N_f \le N_c$ flavors, a dilaton-axion
superfield and a chiral supermultiplet which breaks the 
anomalous $U(1)_X$ symmetry and whose   vacuum expectation value helps
to cancel the $U(1)_X$ D-term.

In section $2$, we analyze in detail this model with global
supersymmetry. It is shown, using
an effective lagrangian approach, that global supersymmetry is
dynamically broken.

In section $3$, we compute the resulting soft breaking terms in
the observable sector and discuss their phenomenological
consequences. We end with some comments.

\section{Gauge group $G=SU(N_c) \times U(1)_X$ with $N_f \le N_c$
flavors}

The model that we consider is an extension of SUSY-QCD based on
the gauge group $SU(N_c)$ with $N_f \le N_c$  flavors of ``quarks''
$Q^i$ of $U(1)_X$ charge $q$ in the fundamental of $SU(N_c)$  and
``antiquarks'' ${\tilde Q}_{\ibar}$ of  charge $\tilde q$ in the
antifundamental of $SU(N_c)$. 

Since we want to avoid $SU(N_c)$ breaking in the $U(1)_X$ flat
direction (\ref{eq:VD}), we  require that the charges $q$ and
$\tilde q$ are positive (this is in fact not restrictive: see 
the comment in the footnote below). We then need at least one field of
negative charge in order to cancel the D-term (\ref{eq:VD}). For
simplicity we will introduce a single field $\phi$ of $U(1)_X$ charge
normalized to $-1$. 

The classical lagrangian compatible with the symmetries is ${\cal L} =
{\cal L}_{kin} + {\cal L}_{couplings}$, where we assume a flat
K\"ahler potential for the matter fields:
\begin{equation}
{\cal L}_{kin} = \int d^4 \theta \left[ Q^+ e^{2q V_X + V_N} Q
+ {\tilde Q} e^{2{\tilde q} V_X - V_N} {\tilde Q}^+ + \phi^+
e^{-2V_X} \phi \right] + {\cal L}_{S,V} \  
\end{equation}
and
\begin{equation}
{\cal L}_{couplings} = \int d^2 \theta ({\phi \over M_P})^{q +
{\tilde q}} m_i^{\ibar} Q^i {\tilde Q}_{\ibar} + h.c. \ .
\label{eq:coupling}
\end{equation}

As stressed in the introduction, the model can be studied {\em per
se} or be used as an illustrative example of a hidden sector where
supersymmetry is dynamically broken in presence of an anomalous
$U(1)$. In the former case, $M_P$ is the scale of the underlying
non-anomalous theory (say the mass of some heavy fermions we have
integrated upon), in the latter case it is the Planck scale.

The mixed anomaly $U(1)_X [SU(N_c)]^2$ which will fix, through
(\ref{eq:Ck}), all the mixed anomalies in the model is given by  
\begin{equation}
C_N = {1 \over 4 \pi^2} N_f (q + {\tilde q}) = k_N \delta_{GS}.
\label{eq:CN} \end{equation} 
We thus require $q + {\tilde q} > 0$, which in turn
justifies the presence of the superpotential term 
(\ref{eq:coupling}).\footnote{Alternatively, since
from (\ref{eq:CN}) $q + \tilde q$ and $\delta_{GS}$ have the same
sign, we would still avoid $SU(N_c)$ breaking in the $U(1)_X$ flat
direction with $q+\tilde q <0$. The field $\phi$ which cancels the
D-term would then be chosen with charge $+1$.}

One may note, using (\ref{eq:shift}), that the following combination
\begin{equation}
f = k_N S - {N_f \over 8 \pi^2} (q+ \tilde q) \ln {\phi \over M_P}
\label{eq:f}
\end{equation}
is invariant under $U(1)_X$. Such a gauge kinetic function would be
obtained by integrating over the hidden matter degrees of freedom, assuming
unbroken supersymmetry. It could then be used to determine the
gaugino masses. We will see however that supersymmetry is broken,
which makes matters less straightforward.

The two scales present in the problem are:

- the scale at which the anomalous $U(1)_X$ symmetry is broken which
is set by
\begin{equation}
\xi = {1 \over 2} k^{1/2}_X g_X \delta^{1/2}_{GS} M_P. \label{eq:ksi}
\end{equation}

- the scale at which the gauge group $SU(N_c)$ enters in a strong
coupling regime:
\begin{equation}
\Lambda = M_P e^{-8\pi^2 k_N S/(3N_c-N_f)}, \label{eq:newLambda}
\end{equation}
where we have used (\ref{eq:Lambda}) with $b_0=(3N_c-N_f)/(16\pi^2)$. 
Notice that, by using the
transformation (\ref{eq:shift}), we find that the dynamical scale
$\Lambda$ has a charge $q_{\Lambda} = N_f (q + {\tilde q})/(3 N_c -
N_f)$.

From now on, we will suppose that $\Lambda << \xi$. We could write
the effective theory below the scale $\xi$ and study within this
theory the strongly coupled $SU(N_c)$ theory. It is however simpler
to keep the complete theory down to the scale $\Lambda$ since most of
the nontrivial effects we will obtain result from an interplay
between the scales $\Lambda$ and $\xi$.

Below the scale $\Lambda$ the appropriate degrees of freedom for
$N_f < N_c$ are the field $\phi$ and the mesons $M_{\ibar}^i = Q^i
{\tilde Q}_{\ibar}$.  The effective superpotential is fixed uniquely
by the global symmetries \cite{TVY,ADS} as follows \begin{equation}
W = (N_c - N_f) {\Lambda^{3N_c - N_f \over N_c - N_f} \over
{(det M)}^{1 \over N_c - N_f}} + ({\phi \over M_P})^{q + 
{\tilde q}} m_i^{\ibar} M_{\ibar}^i \  \label{eq:sup}
\end{equation}
and is seen to be automatically $U(1)_X$ invariant.
Similarly, the gaugino condensation scale 
\begin{equation}
{<\lambda \lambda>}
= \left( \Lambda^{3N_c-N_f}/det M \right)^{1 \over N_c - N_f}
\label{eq:gauginocond}
\end{equation}
is also $U(1)_X$ gauge invariant, as
it should be. 

The gauge contributions to the scalar potential can
be computed along the $SU(N_c)$ classical flat directions. The result
is
\begin{equation}
V_D = {g_X^2 \over 2} \left[ (q + {\tilde q}) Tr (M^+ M)^{1/2}
- \phi^+ \phi + \xi^2 \right]^2 \ . \label{eq:pot}
\end{equation}

The auxiliary fields, computed from (\ref{eq:sup}) and
(\ref{eq:pot}) are
\begin{equation}
{\bar F}_{S^+} = -{8 \pi^2 \over M_P} k_N (S+S^+)^2 {\Lambda^{3N_c -
N_f  \over N_c - N_f} \over {(det M)}^{1 \over N_c - N_f}}
\label{eq:FS} \end{equation}
and
\begin{eqnarray}
{(\bar F_{M^+})}^{\ibar}_i &=&  2  \left[ -{(M^{-1})}_i^{\bar \jmath}
{\Lambda^{3N_c - N_f \over N_c - N_f} \over (det M)^{1 \over N_c -
N_f}} + ({\phi \over M_P})^{q + {\tilde q}} m_i^{\bar \jmath} \right] 
[(M^+ M)^{1/2}]^{\ibar}_{\bar \jmath}, \nonumber \\ 
\bar F_{\phi^+} &=& {q + {\tilde q} \over M_P} ({\phi \over
M_P})^{q + {\tilde q}-1} Tr (m M) \ , \nonumber \\
D_X &=& g_X^2 \left[(q + {\tilde q}) Tr (M^+ M)^{1/2} - \phi^+ \phi +
\xi^2 \right]\ . \label{eq:aux}
\end{eqnarray}

In the limit $S \rightarrow \infty$, the scale $\Lambda$ vanishes
and we can choose $M^i_{\ibar}=0$ and $\phi=\xi$ to cancel all these
auxiliary fields. This is the usual global supersymmetry minimum at
infinite values of $S$ which leads to the dilaton stabilization
problem. We will assume that the dilaton is stabilized at some finite
value $S_0$, possibly through some extra $S$-dependent term in the
superpotential (we will therefore refrain from using (\ref{eq:FS})), 
and that $F_S (S_0) = 0$. Indeed we are going to show that, even in
this unfavorable case (supersymmetry conserving groundstate for $S$),
the other fields present in the theory yield supersymmetry breaking
because of the anomalous behavior of $U(1)_X$. From now on, we will
therefore restrict our attention to the auxiliary fields
(\ref{eq:aux}) associated with $M^i_{\ibar}$, $\phi$ and the $U(1)_X$
gauge degree of freedom.

It is readily seen that the system of equations $F_M = F_{\phi}
= D_X=0$ has no solution as long as $\xi \not= 0$ that is $q +
\tilde q \not= 0$. Therefore supersymmetry is dynamically broken.
As usual, the origin of supersymmetry breaking is chiral: the
non-abelian gauge group content is vector-like but $Q^i$ and $\tilde
Q_{\ibar}$ do not transform in a vectorlike fashion under $U(1)_X$: $q
\not= - \tilde q$; this is precisely what drives the anomaly, which
is therefore the source of chirality in the model.

We will now minimize the scalar potential in terms of $M^i_{\ibar}$,
$\phi$ at fixed $S=S_0$ value such that $F_S(S_0)=0$:
\begin{equation}
V = {1 \over (S+ S^+)^2} {\bar F}_{S^+} F_S + {\bar F}_{\phi^+} F_\phi
+ {1 \over 2} (\bar F_{M^+})^{\ibar}_i [(M^+ M)^{-1/2}]^{\bar
j}_{\ibar}(F_M)^i_{\bar j} + V_D.
\end{equation}
In order to be able to give analytic solutions, we make a few
simplifying asumptions. First, we linearize the minimization
procedure by looking for a minimum in the vicinity of:

a) $\phi_0 = \xi$, the field value which minimizes $V_D$ in the
absence of condensates;

b) $(M_0)^i_{\ibar}$, the solution of $({\bar F}_{M^+})^{\ibar}_i =
0$:
\begin{equation}
(M_0)^i_{\ibar} = (m^{-1})^i_{\ibar} (det m)^{1/N_c} \Lambda^{{3 N_c -
N_f \over N_c}}  \left({\xi \over M_P}\right)^{{N_f - N_c \over
N_c}(q+ \tilde q)} \; \; \; \; .
\end{equation}
One can make a field transformation in order to have a diagonal
matrix  $m^{\ibar}_i$, in which case $(M_0)^i_{\ibar}$ is also
diagonal. Since we are only interested in orders of magnitude, we
will make the assumption that $m^{\ibar}_i = m
\delta^{\ibar}_i$ and search for solutions $M^i_{\ibar} = M
\delta^i_{\ibar}$ of the equations of motion.  

The  minimum is obtained
by making around the field configuration $(M_0)^i_{\ibar} \equiv M_0
\delta^i_{\ibar}$ an expansion in the parameter 
\begin{equation}
\epsilon \equiv {M_0 \over \xi^2} = \left( {\Lambda \over \xi}
\right)^{{3 N_c - N_f \over N_c}} \left[ {m \over M_P} \left({\xi
\over M_P}\right)^{q+ \tilde q-1} \right]^{{N_f - N_c \over N_c}}
\; \; \; . \label{eq:epsilon} 
\end{equation} 
One obtains 
\begin{eqnarray} 
<\phi^+ \phi> &=& \xi^2 \left[ 1 + \epsilon N_f (q+ \tilde q) 
+  \epsilon^2 N_f^2 (q+ \tilde q)^2 \left( -{(N_c -
N_f)(2N_c-2N_f) \over 2 N_c^2} (q+ \tilde q) \right. \right. 
\nonumber \\ 
&+& \left. \left. {1 \over g^2_X} {\hat m}^2  \left[ 1 - {N_f \over
N_c} (q+ \tilde q)\right] \right) + O(\epsilon^3) \right],  \\ 
<M> &=& M_0 \left[ 1 -
{\epsilon \over 2} {N_f (N_c - N_f) (2 N_c-N_f) \over N^2_c} (q+
\tilde q)^2 + O(\epsilon^2) \right], 
\end{eqnarray}
where we have introduced the scale 
\begin{equation}
\hat m = m \left( {\xi \over M_P} \right)^{q+ \tilde q} \ .
\end{equation}  
The value of the auxiliary terms at this ground state are, to leading
order:
\begin{eqnarray}
<D_X> &=& -\epsilon^2 {\hat m}^2 N_f^2 (q+ \tilde q)^2 \left[ 1 -
{N_f \over N_c} (q + \tilde q)\right], \nonumber\\
<F_\phi> &=&   \epsilon {\hat m} \xi N_f (q+ \tilde q), \nonumber \\
<F_M> &=&  -\epsilon^2 {\hat m} \xi^2 {N_f (N_c-N_f) \over N_c} (q +
\tilde q)^2 .
\label{eq:DF} \end{eqnarray}
One may note that $<D_X^{1/2}>$, $<F_\phi / \phi>$ and $<F_M / M>$
are all of the same order $\epsilon \tilde m$. This will have
definite consequences for the soft terms, as we will see in the next
section. Also, using (\ref{eq:epsilon}), one checks that this order
of magnitude goes to zero as $\Lambda \rightarrow 0$, as well as $\xi
\rightarrow \infty$ (at fixed value of $\xi / M_P$). This shows once
again the mixed role that the two scales $\Lambda$ and $\xi$ play as
far as supersymmetry breaking is concerned in this model.

A similar analysis can be performed when $N_f=N_c \equiv N$. In this
case, new degrees of freedom must be introduced in the low energy
effective lagrangian \cite{Seiberg}
\begin{equation}
B =\epsilon_{i_1 \cdots i_N} Q^{i_1} \cdots Q^{i_N}, 
\; \; \tilde B =\epsilon^{{\ibar}_1 \cdots {\ibar}_N} {\tilde
Q}_{{\ibar}_1} \cdots Q_{{\ibar}_N}.
\end{equation}
The effective superpotential compatible with all the symmetries reads
\begin{equation}
W = U \ln {detM - B \tilde B \over \Lambda^{2N}} + \left( {\phi \over
M_P} \right)^{q + \tilde q} m_i^{\ibar} M^i_{\ibar},
\end{equation}
where $U$ is a Lagrange multiplier, physically interpreted as the
gauge composite superfield $U = {\rm Tr} W^{\alpha}_N W_{N \alpha}$.
As in the $N_f < N_c$ case, the system of equations
$F_U=F_M=F_B=F_{\tilde B}=D_X=0$ has no solution and global
supersymmetry is broken.

The properties of the model do not depend either on the assumption of a
single $\phi$ field. Similar conclusions can be reached when one
considers for example a vector-like pair of such fields.

\section{Soft terms in the observable sector}

We now use the results of the preceding section to
determine the order of magnitude of the soft terms in the observable
low energy sector of quarks, leptons and gauginos.

The magnitude of the soft terms in the observable sector is fixed by
the auxiliary fields $F_\phi$, $F_M$ and $D_X$. At the tree level of
the Lagrangian of our model, we find soft scalar masses $\tilde
m_i^2$ and trilinear soft terms $A_{ijk}$ given by the
expressions\footnote{ We assume the presence in the superpotential of
terms of the form $(\phi / M_P)^{X_i + X_j + X_k} \Phi_i \Phi_j
\Phi_k$, as allowed by the $U(1)_X$ symmetry.}: 
\begin{equation} 
\tilde
m_i^2 = X_i  <D_X>, \;\;\; A_{ijk} = (X_i + X_j + X_k) {F_\phi \over
\phi}, \label{eq:mA} \end{equation} where $X_i$ is the $U(1)_X$ charge
of the corresponding field $\Phi^i$. Gaugino masses in the hidden
sector are also induced: $m_\lambda \sim N_f <F_M/M>$. The gaugino
masses in the observable sector are absent at tree level and are
induced by standard gauge loops. Notice that, because supersymmetry is
broken, gaugino masses are not simply given by the gauge invariant
kinetic function (\ref{eq:f}).

In the preceding section we obtained, in the limit $\Lambda \ll \xi$,
the following relation among the auxiliary fields $<F_M / M> \sim
<F_\phi / \phi> \sim <D_X^{1/2}>$. Consequently, all the soft
breaking terms induced at tree level are of the same order. 
By using (\ref{eq:epsilon}) and (\ref{eq:DF}), we obtain
\begin{equation}
\tilde m \sim N_f (q+ \tilde q) {\Lambda^3 
\over \xi^2} \left[  {m \over \Lambda} \left(  {\xi \over M_P}
\right)^{q+ \tilde q} \right]^{N_f / N_c} = N_f (q +
\tilde q) {<\lambda \lambda> \over \xi^2},
\label{eq:softorder}  \end{equation} 
where $\tilde m$ generically denotes a soft-breaking
term (\ref{eq:mA}) and we have used (\ref{eq:gauginocond}) in order to
derive the last relation. This relation is indeed central to the kind
of models described here and stresses the connected role of the
relevant scales: $\xi$ as the scale of messenger interaction and the
gaugino condensate as the seed of supersymmetry breaking (although, as
stressed earlier, the chiral nature of the $U(1)_X$  plays an
important role: $q \not= -\tilde q$).

We now restrict our attention to hidden sector models where the
messengers of supersymmetry breaking are the anomalous $U(1)_X$ gauge
{\em and} gravitational interactions. The scale $M_P$ is therefore the
Planck scale. 
Eq. (\ref{eq:softorder}) should be compared with the
gravitationally-induced soft terms of order 
\begin{equation}
\tilde m|_{grav.} \sim {<\lambda \lambda> \over M_P^2},
\end{equation}
which must be included if the supergravity interactions are switched
on. Since 
\begin{equation}
{1 \over N_f (q + \tilde q)} \left( {\xi \over M_P} \right)^2 
\label{eq:ratio}
\end{equation}
is  a 
small number in the context of superstring models, the supergravity
soft terms can be neglected and the soft terms computed above can be
viewed as phenomenological predictions of these models. It is worth
noting that, using (\ref{eq:CN}) and (\ref{eq:ksi}), the factor $N_f (q
+ \tilde q)$, which may be large and is model dependent, drops out out
of the ratio (\ref{eq:ratio}). One is left with a pure number which
depends on the gauge coupling --as a genuine one-loop effect-- and the
Kac-Moody levels.

If $\xi \sim M_P$, phenomenologically
interesting soft terms $\tilde m \le 1 \ {\rm TeV}$ call for $\Lambda
\ll \Lambda_0$, where $\Lambda_0$ is a typical intermediate
condensation scale: $\Lambda_0 \sim 10^{13} \ {\rm TeV}$.\footnote{
We assume here that $m\sim M_P$.} For $\xi \ll M_P$ the required values
of $\Lambda$ depend on the parameters of the hidden sector $N_f$,
$N_c$, $q + \tilde q$. A generic prediction in this case is a rather
light gravitino: \begin{equation} m_{3/2} \sim  {<W> \over M_P^2} \sim
{N_c \over N_f (q+ \tilde q)} \left( {\xi \over M_P} \right)^2  \tilde
m. \end{equation}
It is useful to notice that generally $N_f (q+ \tilde q) > N_c$.
Then, by using (\ref{eq:DF}), we find $D_X>0$, which means that
scalar particles with positive $U(1)_X$ charges will acquire positive
squared masses. This is a very welcome feature of the model since, by
using the mixed anomaly conditions (\ref{eq:Ck}), (\ref{eq:deltaGS}),
we know that the fields in the observable sector must have
predominantly positive charges. 

This is also consistent with the interpretation of the $U(1)_X$
as an horizontal symmetry which may explain the low energy mass
spectrum \cite{IR,JS,ramond,DPS,Nir}. In this context, positive charges
are   required in the model presented above to account for 
the observed fermion masses and mixings. The anomalous symmetry also
plays an important role in constraining the soft
terms \cite{LNS,DGPS,KK}.

A general feature of the models presented is the large scale
of gauge symmetry breaking. Despite the fact that the messengers of
supersymmetry breaking are mostly gauge interactions,
the resulting soft terms have much lower values, as a
result of a conspiracy between the two scales $\Lambda$ and $\xi$.
This is to be contrasted with the standard gauge-mediated scenarios
where the scale of supersymmetry breaking is much lower
\cite{gaugemed}. In  particular, as we have seen, in the limit where
the two scales are very far away $\Lambda \rightarrow 0$ or $\xi
\rightarrow \infty$, supersymmetry is restored.

There are several aspects which we did not address in this short
paper and which we reserve for further studies: one is the question
of dilaton stabilization since the scalar potential has a complicate
$S$ dependence in its $F$-term through the scale $\Lambda$ and its
$D$-term through the coupling $g_X$. Another is the generalization to
other gauge structures with a chiral content and explicit superstring
realizations.

\vskip 1cm
{\bf Acknowledgments}: We wish to thank  the
Aspen Center for Physics where this work was completed for its
hospitality and the participants of the workshop ``Flavor and Gauge
Hierarchies'', in particular Pierre Ramond, for the lively discussions
and the interesting questions they raised.

\end{document}